\documentclass[
    aps,
    prl,
    twocolumn,
    superscriptaddress,
    groupedaddress,
    nofootinbib
]{revtex4-2}

\usepackage{graphicx}
\usepackage{amsmath}
\usepackage{hyperref}

\newcommand{\bs}{\boldsymbol}

\newcommand{\intk}{\int\!\frac{d^3k}{(2\pi)^3}}
\newcommand{\intkw}{\int\!\frac{d^3k}{(2\pi)^3\,2\omega_k}}

\begin{document}

\title{Interacting Boson System at Finite Temperature:\\
The treatment of the lattice calculations}
\author{D. Anchishkin}
\affiliation{Bogolyubov Institute for Theoretical Physics, 03143 Kyiv, Ukraine}
\author{V. Gnatovskyy}
\affiliation{Bogolyubov Institute for Theoretical Physics, 03143 Kyiv, Ukraine}
\author{D. Zhuravel}
\affiliation{Bogolyubov Institute for Theoretical Physics, 03143 Kyiv, Ukraine}
\author{V. Karpenko}
\affiliation{Bogolyubov Institute for Theoretical Physics, 03143 Kyiv, Ukraine}

\date{\today}

\keywords{Pion gas, phase transition, condensate}

\begin{abstract}
We study interacting relativistic charged bosons at finite
temperature and isospin density in a thermodynamically consistent
mean-field approach with repulsive $\varphi^4$ and $\varphi^6$
interactions.
The thermodynamics is formulated in an Extended Canonical Ensemble, in which
the conserved isospin density, not the chemical potential, is the independent
variable.
This is essential in the condensed phase, where $\mu_I$ is fixed by
condensation.
With one constant fitted to the lattice pressure at $T=120$~MeV, the
model reproduces the lattice isospin density, energy density and
trace anomaly, $\varphi^6$ being more accurate.
\end{abstract}

\maketitle

Hot and dense hadronic matter can be produced in relativistic nucleus-nucleus
collisions, where the densities of thermally excited hadrons can become
sufficiently large for interactions between hadronic degrees of freedom to play
an important role. In this regime, the properties of hadrons are modified by the
surrounding medium, and effective descriptions of the resulting many-body system
become necessary.

A particularly interesting example is a relativistic system of charged bosons
at finite isospin density. At sufficiently large isospin density, Bose
condensation of one of the charged components can occur.
The description of such a system is nontrivial because, in the condensed phase,
the chemical potential reaches the lowest single-particle energy and is
therefore no longer an independent thermodynamic variable. For an interacting
system, the quasiparticle mass is itself a dynamical quantity determined
self-consistently by the scalar density.

In the present work we combine a thermodynamically consistent mean-field
treatment of repulsively interacting relativistic bosons with an Extended
Canonical Ensemble (ECE), in which the conserved isospin density rather than
the isospin chemical potential is taken as the independent variable
(see Refs.~\cite{mishustin-anchishkin-2019,anch-gnat-kondakova-2025}).
We consider $\varphi^4$ and $\varphi^6$ self-interactions and apply the resulting
formalism to a pion-like system.

The main purpose is to test whether this framework can reproduce lattice-QCD
thermodynamics at finite isospin density.
Finite isospin density is one of the few regimes of dense QCD free of the
fermion sign problem and therefore directly accessible to lattice Monte Carlo
simulations~\cite{son-stephanov-2001}.
In particular, we compare the pressure, isospin density, energy density, and
trace anomaly with lattice results at $T = 120$ MeV.

\medskip

We consider a complex scalar field describing charged bosons,
$
{\cal L} = \partial_\mu \hat\phi^\dagger \partial^\mu \hat\phi
- m^2 \hat\phi^\dagger \hat\phi + {\cal L}_{\rm int}
$,
where
\begin{equation}
{\cal L}_{\rm int} = -\frac{\lambda}{2} \left(\hat\phi^\dagger \hat\phi\right)^2,
\qquad
{\cal L}_{\rm int} = -\frac{b}{3} \left(\hat\phi^\dagger \hat\phi\right)^3 \,.
\label{eq:lint}
\end{equation}
%
% with repulsive self-interactions of the forms
The repulsion between pions is what stabilizes the condensate at a finite
isospin density~\cite{son-stephanov-2001}; here this mechanism is implemented
in a thermodynamically consistent mean-field form.

Introducing the scalar density $\sigma=\langle\hat\phi^\dagger\hat\phi\rangle$
and expanding the interaction term around its mean value gives the mean-field
Lagrangian
\begin{equation}
{\cal L}_{\rm MF} = \partial_\mu\hat\phi^\dagger \partial^\mu\hat\phi
- M^2(\sigma)\hat\phi^\dagger \hat\phi + P_{\rm ex}(\sigma),
\label{eq:lmf}
\end{equation}
where
\begin{equation}
M^2(\sigma) = m^2 + U(\sigma),
\qquad
U(\sigma) = -\frac{\partial {\cal L}_{\rm int}}{\partial\sigma},
\label{eq:mass}
\end{equation}
and
\begin{equation}
P_{\rm ex}(\sigma) = {\cal L}_{\rm int}(\sigma)
- \sigma\frac{\partial{\cal L}_{\rm int}}{\partial\sigma}.
\label{eq:pex}
\end{equation}
The quantity $M(\sigma)$ is the effective mass of the quasiparticles, while
$P_{\rm ex}$ represents the interaction contribution to the pressure.
The definitions above satisfy
$\sigma (\partial U/\partial \sigma) = \partial P_{\rm ex}/\partial \sigma$,
which is essential for thermodynamic consistency.
For the two interactions considered here, $U(\sigma) = \lambda \sigma$
for the $\varphi^4$ model and $U(\sigma) = b \sigma^2$ for the $\varphi^6$ model.

In the condensed phase we use the Bogolyubov decomposition \cite{bogolyubov},
$
\hat\phi = \Phi_0 + \hat\psi
$
with
$
\langle\hat\psi\rangle = 0
$,
where $\Phi_0$ describes the condensate and $\hat\psi$ the thermal excitations.
The scalar density is correspondingly separated as
$\sigma = \sigma_{\rm cond} + \sigma_{\rm th}$.
For a homogeneous condensate,
$
\sigma_{\rm cond} = n_{\rm cond}/2M ,
$
where $n_{\rm cond}$ is the condensate particle density.
The quasiparticle dispersion relation is $\omega_k = \sqrt{M^2 + \bs k^2}$.

Throughout we use the compact notation
\begin{equation}
\begin{aligned}
f^{\pm}_k &= \left[ e^{(\omega_k \mp \mu_I)/T} - 1 \right]^{-1} , \\[2pt]
L^{\pm}_k &= \ln\!\left[ 1 - e^{-(\omega_k \mp \mu_I)/T} \right] ,
\end{aligned}
\label{eq:notation}
\end{equation}
so that $f^{+}_k$ and $f^{-}_k$ refer to particles and antiparticles,
respectively.
The conserved isospin density is obtained from the Noether current.
In the condensed phase it has the form
$n_I = n_{\rm cond} + \int d^3k/(2\pi)^3 \left( f^{+}_k - f^{-}_k \right)$.
Because the two charged components would require simultaneously
$M - \mu_I = 0$ and $M + \mu_I = 0$
in order to condense, simultaneous condensation is impossible for a massive
repulsively interacting system\footnote{If a system, in addition to repulsive
interactions, possesses a strong attractive interaction, it becomes possible.}.
At nonzero isospin density only one charged component can therefore form a
condensate.
We take this component to be the positively charged pion-like state.
Therefore, in the condensed phase $\mu_I$ is to be replaced by $M(\sigma)$
everywhere in Eq.~(\ref{eq:notation}).

\medskip

The thermodynamic description starts from the grand canonical ensemble,
$
Z(T,\mu_I,V) = {\rm Tr}\, \exp[-\beta(H_{\rm MF}-\mu_I\hat N_I)] ,
$
with the grand potential $\Omega = -T\ln Z$. In the thermal and in the
condensed phase, respectively,
\begin{eqnarray}
\Omega &=& \Omega_{\rm th} - VP_{\rm ex} \,,
\label{eq:omega-th}
\\
\Omega &=& N_{\rm cond}\left( M - \mu_I \right)
+ \Omega_{\rm th} - VP_{\rm ex} \,,
\label{eq:omega-cond}
\end{eqnarray}
where
\begin{equation}
\Omega_{\rm th} = VT \intk \left( L^{+}_k + L^{-}_k \right) .
\label{eq:omega-thermal}
\end{equation}
The condensate term in Eq.~(\ref{eq:omega-cond}) vanishes identically once the
onset condition $\mu_I = M$ is imposed, but it has to be retained in the
variational procedure that determines $n_{\rm cond}$.
The isospin density is $n_I = -V^{-1}\partial\Omega/\partial\mu_I$.
Rather than regarding $\mu_I$ as the independent variable, we solve this
relation for $\mu_I = \mu_I(T,n_I)$ and perform the Legendre transformation
\begin{equation}
F(T,N_I,V) = \Omega - \mu_I \frac{\partial\Omega}{\partial\mu_I} \,.
\label{eq:legendre}
\end{equation}
The free-energy density $\mathcal{F} = F/V$ therefore becomes a function of the
canonical variables $T$ and $n_I$.
This construction is especially important in the condensed phase.
There the chemical potential is constrained by the onset condition
\begin{equation}
% \boxed{\;\mu_I = M(\sigma)\;}
\mu_I \,=\, M(\sigma) \,,
\label{eq:onset}
\end{equation}
and hence cannot be varied independently of the thermodynamic state.
The effective mass must instead be obtained together with the scalar and
condensate densities from a self-consistent set of equations.

In the thermal phase these equations are
\begin{eqnarray}
n_I &=& \intk \left( f^{+}_k - f^{-}_k \right) ,
\label{eq:isospin-dens-1}
\\
\sigma &=& \intkw \left( f^{+}_k + f^{-}_k \right) .
\label{eq:sigma-1}
\end{eqnarray}
%
% with $M^2 = m^2 + U(\sigma)$.
In the condensed phase, where $\mu_I = M$, they become
\begin{eqnarray}
n_I &=& n_{\rm cond} + \intk \left( f^{+}_k - f^{-}_k \right) ,
\label{eq:isospin-dens-2}
\\
\sigma &=& \frac{n_{\rm cond}}{2M}
+ \intkw \left( f^{+}_k + f^{-}_k \right) ,
\label{eq:sigma-2}
\end{eqnarray}
again together with $M^2 = m^2 + U(\sigma)$.
These equations determine $\sigma(T,n_I)$, $M(T,n_I)$, $n_{\rm cond}(T,n_I)$
without treating $\mu_I$ as an independent variable in the condensed phase.
In the ECE the pressure follows from $p = n_I \mu_I - \mathcal{F}$, and can be
written as
\begin{equation}
p = -T \intk \left( L^{+}_k + L^{-}_k \right) + P_{\rm ex} \,,
\label{eq:pressure}
\end{equation}
where $\mu_I = \mu_I(T,n_I)$.
The first term in Eq.~(\ref{eq:pressure}) is the kinetic pressure of the thermal
quasiparticles and antiparticles, while $P_{\rm ex}$ contains the interaction
contribution.
The condensate itself does not produce an independent kinetic-pressure
contribution.
The energy density is obtained thermodynamically from
$
\epsilon = \mathcal{F} - T \left(\partial \mathcal{F}/\partial T\right)_{n_I} .
$
In the condensed phase this gives
\begin{equation}
\epsilon = Mn_{\rm cond}
+ \intk \omega_k \left( f^{+}_k + f^{-}_k \right) - P_{\rm ex} \,.
\label{eq:energy-density-2}
\end{equation}

\medskip

For comparison with finite-isospin lattice QCD it is necessary to
specify carefully the normalization of the isospin chemical
potential.
For a pion system, $N_I = N_{\pi^+}-N_{\pi^-}$. Using the
quark content $\pi^+ = u\bar d$, $\pi^- = d\bar u$ and the corresponding
quark numbers, the pion isospin number can be written as
$N_I=N_u-N_d$.
If the quark isospin chemical potential is defined
through $\mu_I^{(q)} = \mu_u-\mu_d$, then consistency of the pion and
quark descriptions gives $\mu_I^{(\pi)} = \mu_I^{(q)}$.
With this normalization the condensation condition for an ideal pion gas is
$\mu_I = m_\pi$.
For example,
this is the convention of Son and Stephanov~\cite{son-stephanov-2001},
who assign to the light quarks chemical potentials of equal magnitude
$|\mu_I|/2$ and opposite sign, and find the onset of pion condensation at
$|\mu_I| = m_\pi$ (see also \cite{anch-gnat-kondakova-2025} and references
therein).
The lattice calculations
\cite{brandt-2021,brandt-2022,brandt-2018,brandt-2016,brandt-2017}
considered here instead employ the normalization
$\mu_I=\frac{1}{2}(\mu_u-\mu_d)$, while retaining $N_I = N_u - N_d$.
With this convention the pion chemical potential entering the Bose distribution
is $2\mu_I$, and the condensation condition becomes
\begin{equation}
% \boxed{\;\mu_I=\frac{m_\pi}{2}\;}.
\mu_I \,=\, \frac{m_\pi}{2} \,.
\label{eq:threshold}
\end{equation}
Thus the difference between the two thresholds is a normalization of the isospin
chemical potential.
For the comparison below we use the lattice convention.

%1
\begin{figure*}[t]
\includegraphics[width=0.43\textwidth]{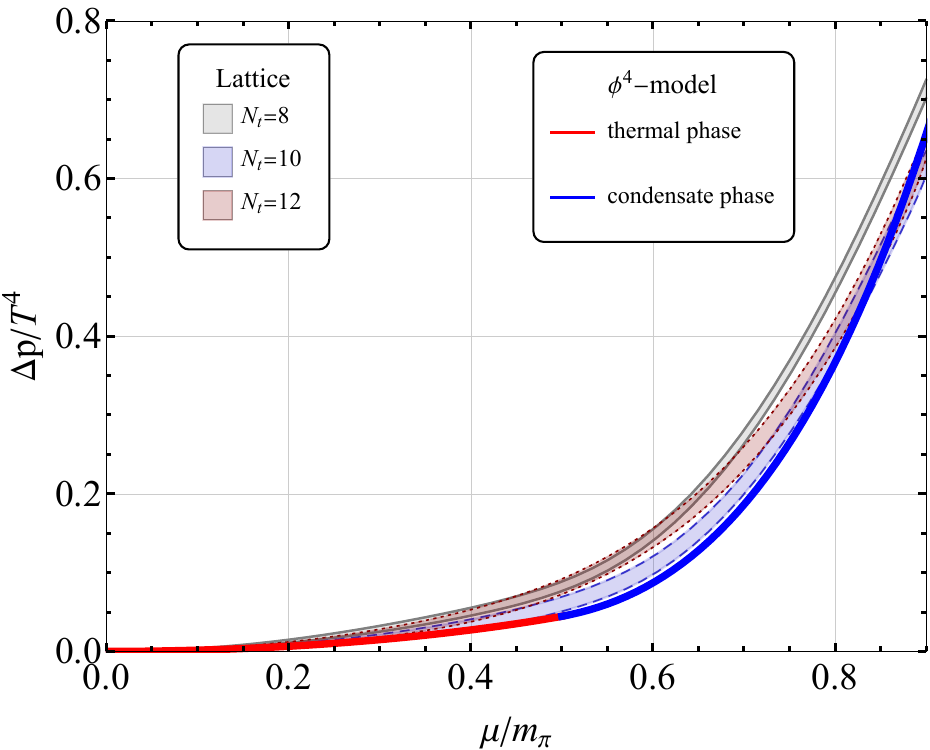}
\includegraphics[width=0.43\textwidth]{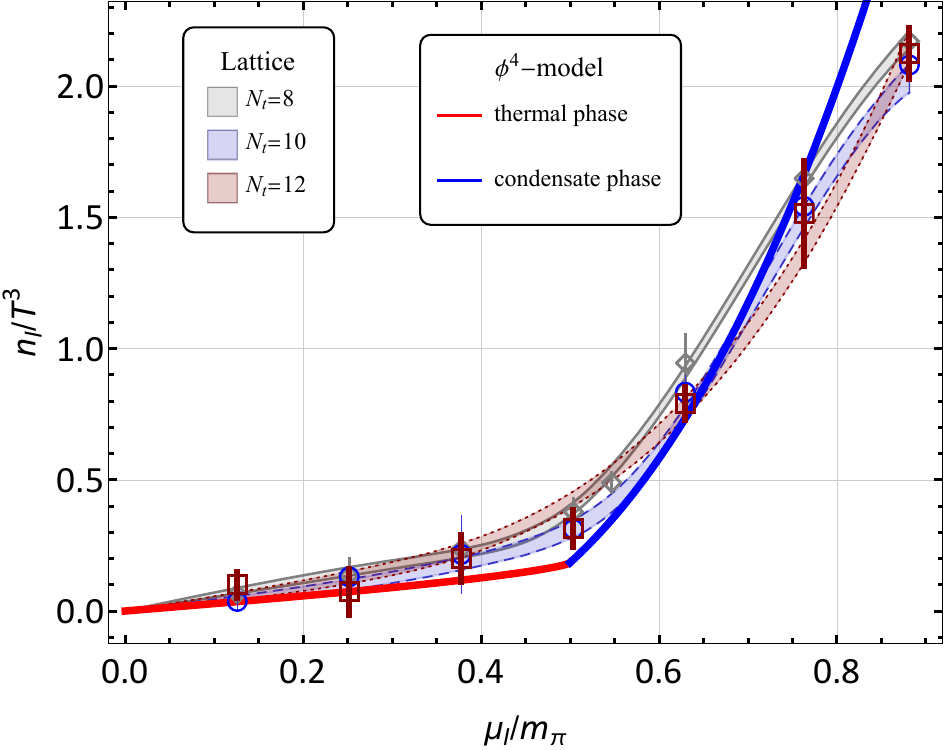}
\includegraphics[width=0.43\textwidth]{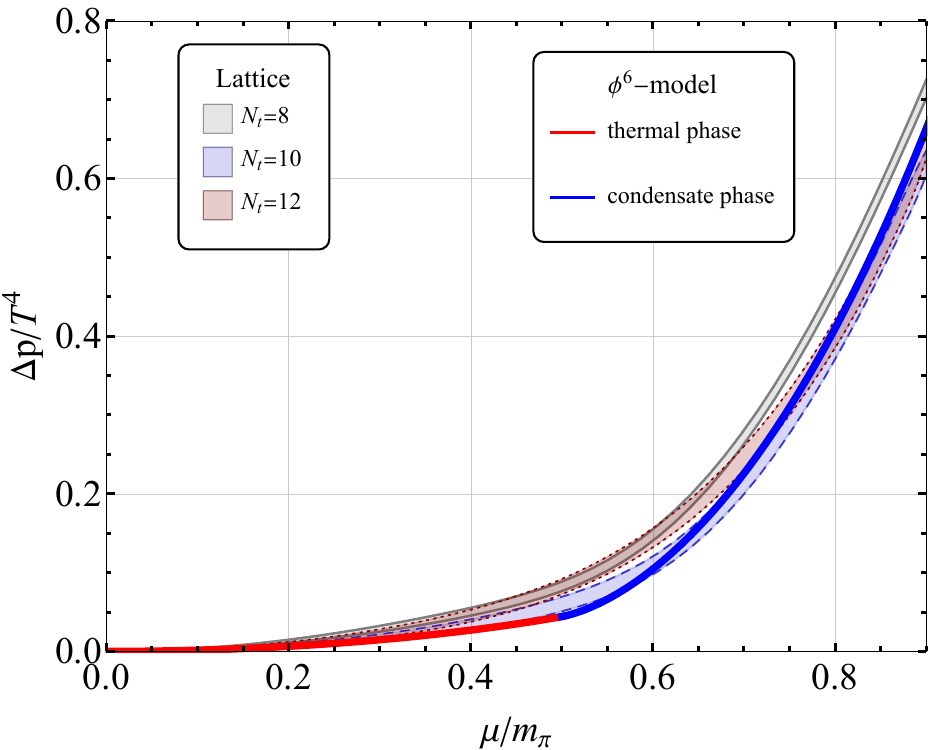}
\includegraphics[width=0.43\textwidth]{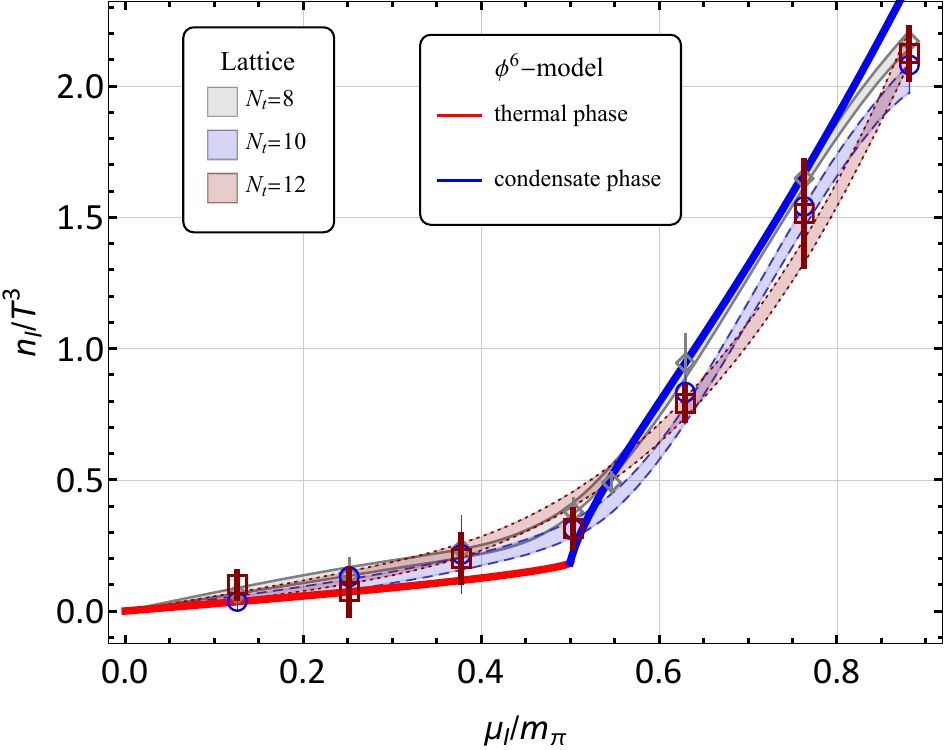}
\caption{\small
Fit of lattice calculations results \cite{brandt-2022} at $T = 120~$MeV for
the quantity $\Delta p$ in the $\varphi^4$ model (top left panel) and
the $\varphi^6$ model (bottom left panel).
Description of lattice results \cite{brandt-2022} at $T = 122~$MeV
for $n_I$ in the $\varphi^4$ model (top right panel) and the $\varphi^6$ model
(bottom right panel).
The red segment of the approximation curve refers to the thermal phase, and
the blue segment to the condensed phase.
 }
\label{fig:P-T122}
\end{figure*}

We now apply the interacting-boson formalism to the lattice-QCD results for
finite isospin density \cite{brandt-2021,brandt-2022,brandt-2018}.
The temperature is fixed at $T=120$~MeV, and the pion mass used in the lattice
analysis is $m_\pi = 135$~MeV.
In the adopted convention the onset of the condensed phase occurs at
$\mu_I = m_\pi/2$.
Since the condensate condition in our model is $\mu_I = M$, the effective
quasiparticle mass at the critical point must satisfy
$M(\sigma_{\rm cr}) = m_\pi/2$.
The critical scalar density is consequently determined from
(\ref{eq:sigma-2}) at $\sigma_{\rm cond} = 0$,
\begin{equation}
\sigma_{\rm cr} = \intkw \left( f^{+}_k + f^{-}_k \right)
\bigg|_{\mu_I = m_\pi/2} \,,
\label{eq:sigma-cr}
\end{equation}
with $\omega_k = \sqrt{m_\pi^2/4 + \bs k^2}$.
For the $\varphi^4$ and the $\varphi^6$ model, respectively,
\begin{equation}
M^2 = m_*^2 + \lambda \sigma_{\rm cr},
\qquad
M^2 = m_*^2 + b \sigma_{\rm cr}^2 .
\label{eq:msq}
\end{equation}
The interaction parameters are determined by fitting the pressure.
The resulting values are
\begin{equation}
\lambda \simeq 0.4934, \qquad  b m_\pi^2 \simeq 0.5673 .
\label{eq:interac-constants}
\end{equation}
The corresponding fitted bare masses are $m_* = m_\pi/2.22$ for the $\varphi^4$
model and $m_* = m_\pi/2.021$ for the $\varphi^6$ model.

%2
\begin{figure*}[t]
\includegraphics[width=0.43\textwidth]{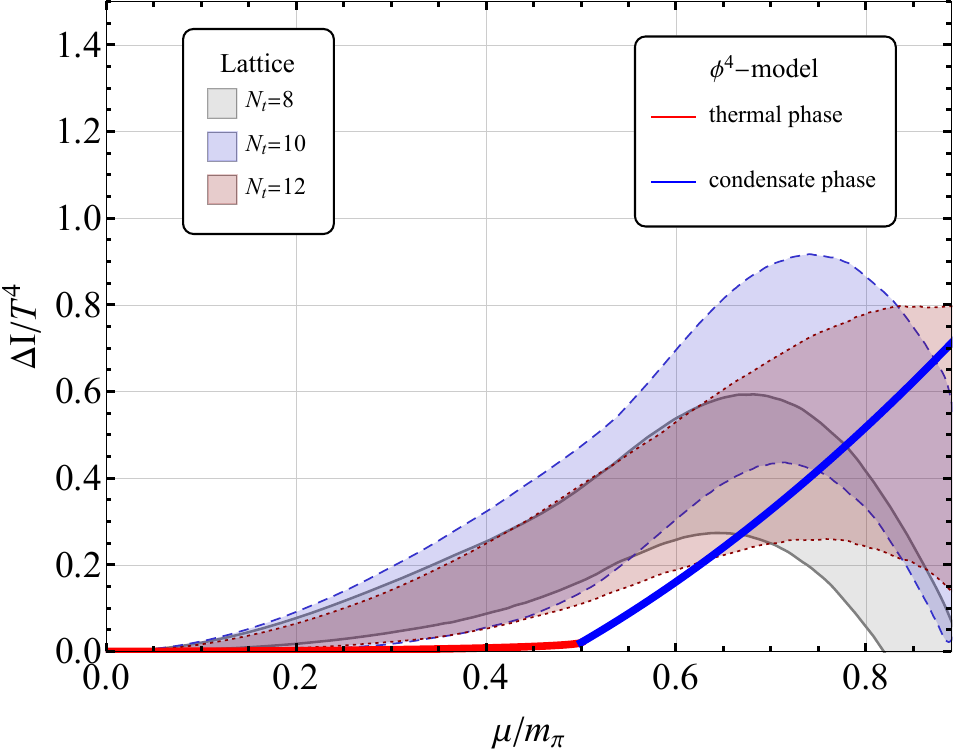}
\includegraphics[width=0.43\textwidth]{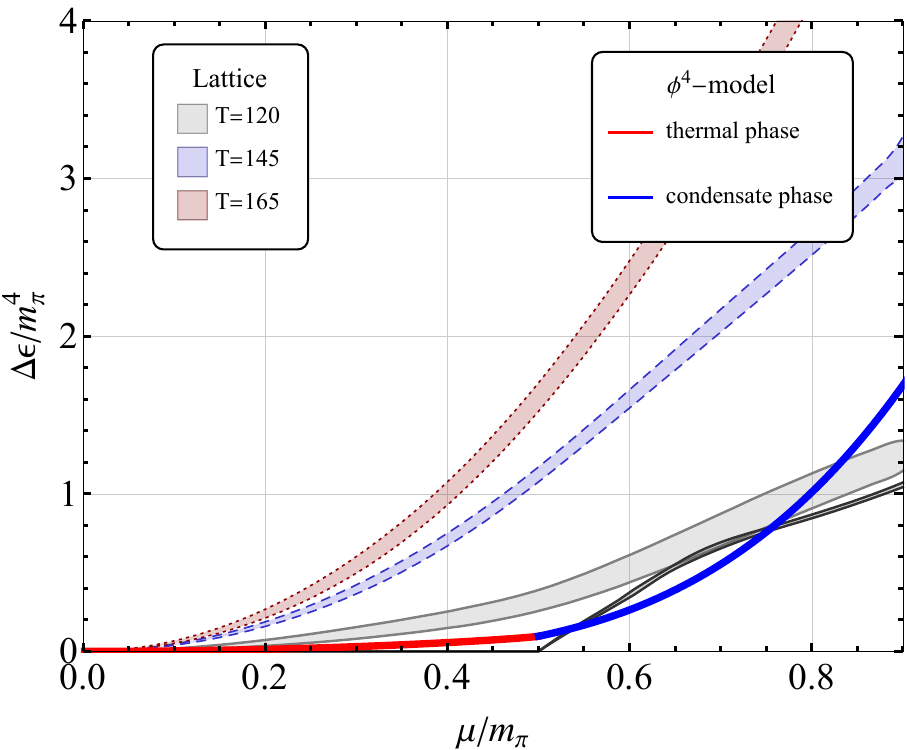}
\includegraphics[width=0.43\textwidth]{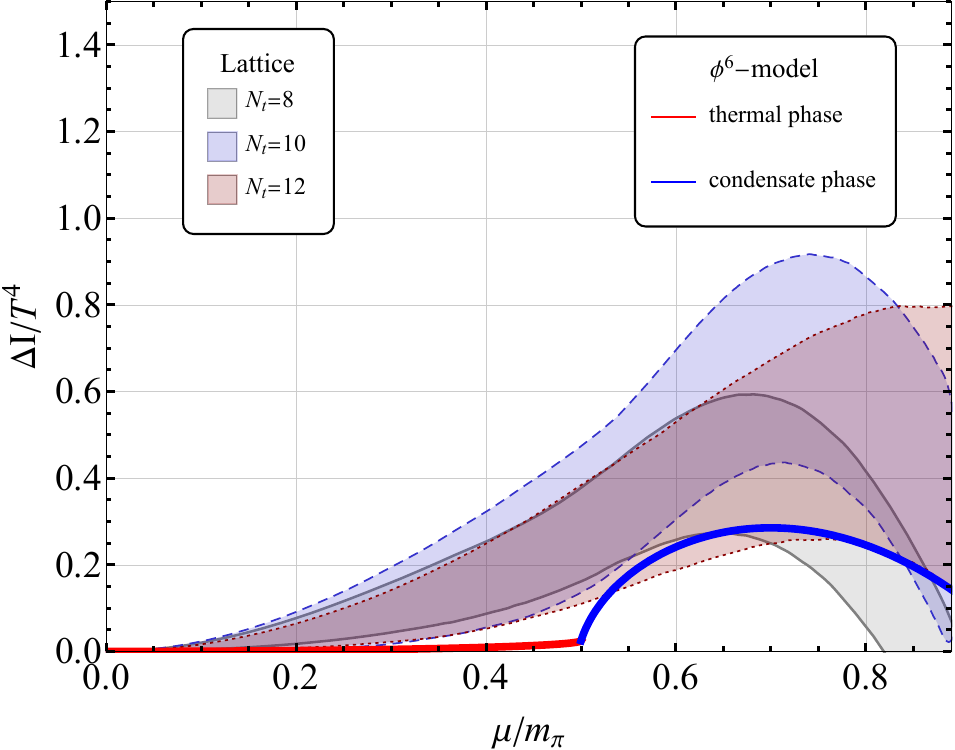}
\includegraphics[width=0.43\textwidth]{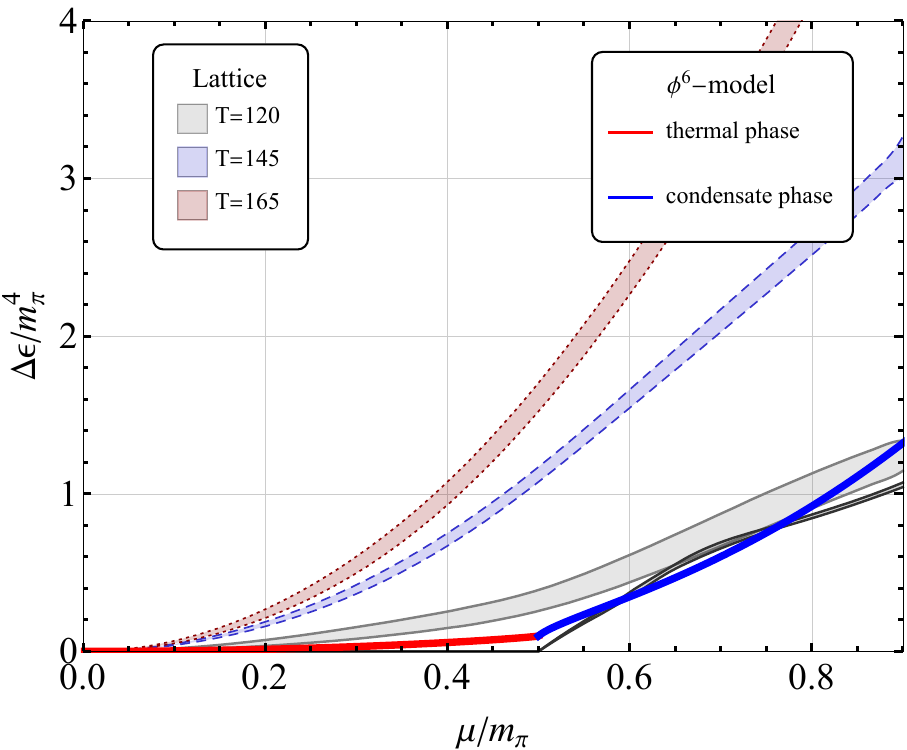}
\caption{\small Description of lattice calculation results
\cite{brandt-2021,brandt-2022} at $T = 120~$MeV
for the quantities $\Delta I$  and $\Delta \varepsilon$ in the $\varphi^4$
model (top panels) and the $\varphi^6$ model (bottom panels).
The red segment of the fit curve corresponds to the thermal phase, and
the blue segment to the condensed phase.
 }
\label{fig:delta-I-T122}
\end{figure*}

The comparison with lattice data is shown in
Figs.~\ref{fig:P-T122} and \ref{fig:delta-I-T122}, the theoretical
approximation curve in each panel consists of two segments:
the red segment corresponds to the thermal phase, and the blue segment
to the condensed phase.
The point where these two segments meet corresponds to the phase transition
in the creation of the Bose-Einstein condensate.
In the thermal region, the chemical potential is obtained by solving the
thermal self-consistency equations at fixed $T$ and $n_I$.
In the condensed region it is constrained by $\mu_I = M(T,n_I)$,
with the effective mass determined from the coupled condensate equations.
Thus, although the horizontal axis of the comparison
plots is labeled $\mu_I$, the calculation in the condensed region
uses the conserved density as the independent thermodynamic
variable, and the plotted value of the chemical potential is then
obtained as $\mu_I = M(\sigma)$.
In the thermal phase, the plotted value of $\mu_I$ is obtained as the solution
of Eqs.~(\ref{eq:isospin-dens-1}) and (\ref{eq:sigma-1}).
This is the essential role of the Extended Canonical Ensemble in the present
analysis.

Both interaction models reproduce the main behavior of the lattice pressure
($p = p(T,0) + \Delta p(T,\mu_I)$) and isospin density.
The $\varphi^6$ model gives a visibly better overall description of the lattice
results.
In Fig.~\ref{fig:delta-I-T122} the same comparison is extended to the energy
density ($\varepsilon = \varepsilon(T,0) + \Delta \varepsilon(T,\mu_I)$) and
trace anomaly ($I = \varepsilon - 3p = I(T,0) + \Delta I(T,\mu_I)$).

The agreement is nontrivial because the quantities shown in the different panels
are not independently fitted: once the interaction parameters $\lambda$ or $b$
are fixed from the
pressure, the isospin density, energy density, and trace anomaly follow from the
same thermodynamically consistent framework.

The comparison therefore provides a test of both the interaction model and the
thermodynamic formulation.
In particular, the continuous matching of the thermal and condensate branches
demonstrates the practical advantage of treating the conserved isospin density
as the canonical variable when the chemical potential becomes constrained
by condensation.

\medskip

We have developed a thermodynamically consistent mean-field description of an
interacting relativistic particle-antiparticle boson system at finite
temperature and fixed isospin density.
Repulsive $\phi^4$ and $\phi^6$ interactions generate a density-dependent
quasiparticle mass and an associated excess-pressure contribution.
The resulting thermodynamics is formulated in the Extended Canonical Ensemble,
obtained through a Legendre transformation from the isospin chemical potential
to the conserved isospin density.

The ECE formulation is particularly useful in the Bose-condensed phase, where
the chemical potential is constrained by $\mu_I = M$ and therefore cannot be
regarded as an independent thermodynamic variable.
The condensate and effective mass are instead obtained self-consistently at
fixed temperature and isospin density.

Applied to a pion-like system, the framework reproduces the main features of
finite-isospin lattice-QCD thermodynamics at $T = 120$~MeV.
After applying the isospin-chemical potential normalization used in the
lattice calculations, both models, $\varphi^4$ and $\varphi^6$, after pressure
correction with only one fitting parameter (the interaction constant), give
reasonable descriptions of the pressure, isospin density, energy density,
and trace anomaly.
Among the two, the $\varphi^6$ interaction provides the better overall description.
At the same time, the success of the presented interacting pion-like models
suggests that, at ``hadron'' temperatures, at least at $T = 120$~MeV,
and at the considered isospin densities, the vast majority of quark
configurations reflect pion structure,
with $n_{\bar d} = n_u$ and $n_{\bar u} = n_d$.

The results indicate that an Extended Canonical formulation provides a useful
framework for studying interacting relativistic bosonic matter across the
transition between the thermal and Bose-condensed regimes and for connecting
effective bosonic descriptions with lattice-QCD thermodynamics at finite isospin
density.

\medskip

\begin{acknowledgments}
This work was supported by the National Research Foundation of Ukraine under
Grant No. 2025.07/0461.
Authors thank A.~Korchin for useful and fruitful discussions.
\end{acknowledgments}

%%%%%%%%%%%%%%%%%%%%%%%%%%%


\begin{thebibliography}{40}

\bibitem{mishustin-anchishkin-2019}
    I.N. Mishustin, D.V. Anchishkin, L.M. Satarov, O.S. Stashko, and H. Stoecker,
    Condensation of interacting scalar bosons at finite temperatures,
    Phys. Rev. C {\bf 100}, 022201(R) (2019);
\newline [\href{https://doi.org/10.1103/PhysRevC.100.022201}{DOI: 10.1103/PhysRevC.100.022201}];
\newline [\href{https://arxiv.org/abs/1905.09567}{arXiv: 1905.09567 [nucl-th]}].
%    [DOI: https://doi.org/10.1103/PhysRevC.100.022201];
%    arXiv:1905.09567 [nucl-th] (2019).

\bibitem{anch-gnat-kondakova-2025}
D. Anchishkin, V. Gnatovskyy, and I. Kondakova,
Ideal Boson Particle-Antiparticle System at Finite Temperatures,
J. Phys. G: Nucl. Part. Phys. {\bf 52}, 115002 (2025);
\newline [\href{https://doi.org/10.1088/1361-6471/ae1152}{DOI: 10.1088/1361-6471/ae1152}];
\newline [\href{https://arxiv.org/abs/2507.10752}{arXiv: 2507.10752 [nucl-th]}].
% [DOI 10.1088/1361-6471/ae1152];
% arXiv: 2507.10752 [nucl-th] (2025).

\bibitem{son-stephanov-2001}
D.T. Son and M.A. Stephanov, QCD at finite isospin density,
Phys. Rev. Lett. {\bf 86}, 592 (2001);
\newline [\href{https://doi.org/10.1103/PhysRevLett.86.592}{DOI: 10.1103/PhysRevLett.86.592}];
\newline [\href{https://arxiv.org/abs/hep-ph/0005225}{arXiv: hep-ph/0005225 [hep-ph]}].
% QCD at a finite isospin density: From the pion to quark-antiquark condensation,
% Phys. Atom. Nucl. {\bf 64}, 834 (2001) [arXiv:hep-ph/0011365].

\bibitem{bogolyubov}
N. Bogolubov,
On the theory of superfluidity, Sov. J. Phys. {\bf 11}, 23 (1947).
%
M.M. Bogolyubov, {\it Lekciyi z kvantovoyi statystyky},
Kyiv, 1947 (Ukrainian).
%
N.N. Bogoliubov, {\it Lectures on Quantum Statistics},
Gordon and Breach, New York, 1967.

\bibitem{brandt-2021}
Bastian B. Brandt, Francesca Cuteri and Gergely Endr\H{o}di,
QCD thermodynamics at non-zero isospin asymmetry,
POS, LATTICE2021 {\bf 132}  (2022).
\newline [\href{https://doi.org/10.22323/1.396.0132}{DOI: 10.22323/1.396.0132}];
\newline [\href{https://arxiv.org/abs/2110.14750}{arXiv: 2110.14750 [hep-lat]}].

\bibitem{brandt-2022}
B.B. Brandt, F. Cuteri, and  G. Endr\H{o}di,
Equation of state and speed of sound of isospin-asymmetric QCD on the lattice,
JHEP {\bf 07}, 055 (2023);
\newline [\href{https://doi.org/10.1007/JHEP07\%282023\%29055}{DOI: 10.1007/JHEP07(2023)055}];
\newline [\href{https://arxiv.org/abs/2212.14016}{arXiv: 2212.14016 [hep-lat]}].
% DOI:10.1007/JHEP07(2023)055; arXiv:2212.14016 [hep-lat].
% Published: Jul 6, 2023

\bibitem{brandt-2018}
B.B. Brandt, G. Endr\H{o}di and S. Schmalzbauer,
QCD phase diagram for nonzero isospin-asymmetry,
Phys. Rev. D {\bf 97}, 054514  (2018);
\newline [\href{https://doi.org/10.1103/PhysRevD.97.054514}{DOI: 10.1103/PhysRevD.97.054514}]
\newline [\href{https://arxiv.org/abs/1712.08190}{arXiv: 1712.08190 [hep-lat]}].

\bibitem{brandt-2016}
B.B. Brandt, G. Endr\H{o}di,
QCD phase diagram with isospin chemical potential (14p.),
PoS LATTICE2016 {\bf 039} (2016);
\newline [\href{https://doi.org/10.22323/1.256.0039}{DOI: 10.22323/1.256.0039}];
\newline [\href{https://arxiv.org/abs/1611.06758}{arXiv: 1611.06758 [hep-lat]}].
% See Section: Introduction --> $\mu_I = (\mu_u - \mu_d)/2$,  eq.(1.1)

\bibitem{brandt-2017}
B.B. Brandt, G. Endr\H{o}di, S. Schmalzbauer,
QCD at finite isospin chemical potential,
EPJ Web Conf., {\bf 175}, 07020 (2018);
% https://doi.org/10.1051/epjconf/201817507020;
\newline [\href{https://doi.org/10.1051/epjconf/201817507020}{DOI: 10.1051/epjconf/201817507020}];
\newline [\href{https://arxiv.org/abs/1709.10487}{arXiv: 1709.10487 [hep-lat]}].
%Bastian B. Brandt, Gergely Endrodi, Sebastian Schmalzbauer (Frankfurt U.)(16p)
% See Section: Introduction --> $\mu_I = (\mu_u - \mu_d)/2$

\end{thebibliography}
\end{document}